\documentclass[%
 reprint,
 superscriptaddress,
 amsmath,amssymb,
 aps,
]{revtex4-2}
\usepackage[english]{babel}
\usepackage{microtype}
\usepackage[utf8]{inputenc}

\usepackage{graphicx}
\usepackage{dcolumn}
\usepackage{bm}
\usepackage{natmove}

\usepackage{zebra-goodies}

\begin{document}

\title{Scale-dependent roughness parameters for topography analysis}

\author{Antoine Sanner}
\affiliation{Department of Microsystems Engineering, University of Freiburg, Georges-K\"ohler-Allee 103, 79110 Freiburg, Germany}
\affiliation{Cluster of Excellence livMatS, Freiburg Center for Interactive Materials and Bioinspired Technologies, University of Freiburg, Georges-K\"ohler-Allee 105, 79110 Freiburg, Germany}

\author{Wolfram G. Nöhring}
\affiliation{Department of Microsystems Engineering, University of Freiburg, Georges-K\"ohler-Allee 103, 79110 Freiburg, Germany}

\author{Luke A. Thimons}
\affiliation{Department of Mechanical Engineering and Materials Science, University of Pittsburgh, 3700 O’Hara Street, Pittsburgh, Pennsylvania 15261, USA}

\author{Tevis D. B. Jacobs}
\affiliation{Department of Mechanical Engineering and Materials Science, University of Pittsburgh, 3700 O’Hara Street, Pittsburgh, Pennsylvania 15261, USA}

\author{Lars Pastewka}
\email{lars.pastewka@imtek.uni-freiburg.de}
\affiliation{Department of Microsystems Engineering, University of Freiburg, Georges-K\"ohler-Allee 103, 79110 Freiburg, Germany}
\affiliation{Cluster of Excellence livMatS, Freiburg Center for Interactive Materials and Bioinspired Technologies, University of Freiburg, Georges-K\"ohler-Allee 105, 79110 Freiburg, Germany}

\date{\today}

\begin{abstract}
The failure of roughness parameters to predict surface properties stems from their inherent scale-dependence; in other words, the measured value depends on the way it was measured. Here we take advantage of this scale-dependence to develop a new framework for characterizing rough surfaces: the Scale-Dependent Roughness Parameters (SDRP) analysis that yields slope, curvature and higher-order derivatives of surface topography at many scales, even on a single topography measurement. We demonstrate the relationship between SDRP and other common statistical methods for analyzing surfaces: the height-difference autocorrelation function (ACF), variable bandwidth methods (VBMs) and the power spectral density (PSD). We use computer-generated and measured topographies to demonstrate the benefits of SDRP analysis, including: novel metrics for characterizing surfaces across scales, and the detection of measurement artifacts. The SDRP is a generalized framework for scale-dependent analysis of surface topography that yields metrics that are intuitively understandable.
\end{abstract}

\maketitle

\section{Introduction}

Surface roughness is primarily characterized in terms of scalar parameters; especially common are the root-mean-square (rms) height and slope which are the rms deviations from the respective mean values. Some variant of these quantities is computed by all surface topography instruments and they are often reported to describe surface topography in publications. They quantify the amplitude of spatial fluctuations in height and slope across the measured topography. A core issue with these roughness parameters is that all of them explicitly depend on the scale of the measurement~\cite{jacobs_quantitative_2017}: The rms height depends on the lateral size (largest scale) of the measurement; the rms slope depends on the resolution (smallest scale) of the measurement. A direct demonstration of this effect on real-world measurements can be found in Refs.~\cite{gujrati_combining_2018,gujrati_comprehensive_2021}.

This scale-dependence is typically a signature of the multi-scale nature of surface topography. A simple illustration is given in a classic article by Benoit Mandelbrot on the length of coastlines~\cite{mandelbrot_how_1967}. Mandelbrot illustrated, that the length $L$ of a coastline depends on the length of the yardstick $\ell$ used to measure it. A smaller yardstick picks up finer details and hence leads to longer coastlines. For (self-affine) fractals~\cite{mandelbrot_fractal_1982}, the functional relationship $L(\ell)$ is a power-law whose exponent characterizes the fractal dimension of the coastline. In the case of a surface topography measurement, $\ell$ corresponds to the resolution of the scientific instrument used to measure the topography and the property corresponding to the length of a coastline is the true surface area $A(\ell)$ of the topography. We have in prior work directly demonstrated that $A(\ell)$ (and also the rms slope and curvature) scales with measurement resolution $\ell$~\cite{gujrati_combining_2018,dalvi_linking_2019,gujrati_comprehensive_2021}. This scaling of the surface area has, for example, direct relevance to adhesion between soft surfaces~\cite{dalvi_linking_2019}. While many surfaces may not be ideal fractals, they still exhibit some form of size dependence of the roughness parameters discussed above. This is because processes that shape surfaces, such as fracture~\cite{mandelbrot_fractal_1984,ponson_two-dimensional_2006,bonamy_scaling_2006}, plasticity~\cite{nadgorny_evolution_2006,schwerdtfeger_scale-free_2007,sandfeld_deformation_2014,hinkle_emergence_2020} or erosion~\cite{persson_fractal_2014}, all lead to fractal scaling over a range of length scales.

We here suggest a route to generalize these (and other) geometric properties of measured topography to explicitly contain a notion of measurement scale. We define the individual roughness parameter as a function of scale $\ell$ over which it is measured, leading to curves identifying the parameter as a function of $\ell$. We term the resulting curves the \emph{scale-dependent roughness parameters (SDRPs)} and outline the relationship to three common characterization techniques: the height-difference autocorrelation function (ACF), the variable bandwidth method (VBM) and the power spectral density (PSD). SDRPs are useful because they are easily interpreted: While it is difficult to attach a geometric meaning to a certain value of the PSD (where even units can be unclear~\cite{jacobs_quantitative_2017}), slope and curvature both have simple geometric interpretations. Since slope and curvature are also the primary ingredients for modern theories of contact between rough surfaces~\cite{bush_elastic_1975,persson_elastoplastic_2001,persson_theory_2001,persson_effect_2001,persson_adhesion_2002,hyun_finite-element_2004,campana_contact_2007,muser_rigorous_2008,pastewka_contact_2014}, SDRPs are directly connected to functional properties of rough surfaces. Finally, we illustrate below how SDRPs can be used to estimate tip radius artifacts in scanning probe measurements.

\section{Analysis methodology}

\subsection{Computing typical roughness parameters in real space}

Surface topography is commonly described by a function $h(x,y)$, where $x$ and $y$ are the coordinates in the plane of the surface. This is sometimes called the Monge representation of a surface, which is an approximation as it excludes overhangs (reentrant surfaces). 
A real measurement does not yield a continuous function but height values
\begin{equation}
    h_{kl}=h(x_k, y_l)
    \label{eq:collocation}
\end{equation}
on a set of discrete points $x_k$ and $y_l$.
Measurements are often taken on equidistant samples where $x_k = k\Delta x$ and $y_l = l\Delta y$, where $\Delta x$ and $\Delta y$ is the difference between the sample points. Furthermore $k\in[0, N_x-1]$ and $l\in[0, N_y-1]$ where $N_x\times N_y$ is the total number of sample points. 

Topographies are often random such that $h_{kl}$ is a random process and its properties must be described in a statistical manner. Starting with Longuet-Higgins~\cite{longuet-higgins_statistical_1957,longuet-higgins_statistical_1957-1} and Nayak~\cite{nayak_random_1971}, many authors have discussed this random process model of surface roughness but the most commonly used roughness parameters have remained simple.

We will illustrate the following concepts using the one-dimensional case, i.e.\ for line scans or profiles. In many real scenarios, even topographic data is interpreted as a series of line scans. This is for example the case in atomic force microscopy (AFM), where a topographic map is stitched together from adjacent line scans. Because of temporal (instrumental) drift these line scans may not be perfectly aligned and the ``scan''-direction is then the preferred direction for statistical evaluation. In the following mathematical development, we will implicitly assume that all values are obtained by averaging over these consecutive scans, but we will not write this average explicitly in the equations that follow. Extension to true two-dimensional topography maps of the ideas presented here is straightforward and briefly discussed in Appendix~\ref{app:2d}.

The most straightforward statistical property is the root-mean-square (rms) height,
\begin{equation}
    h_\text{rms} = \left\langle h_k^2 \right\rangle^{1/2} \equiv \left\langle h^2(x) \right\rangle^{1/2}, 
\end{equation}
where the average $\langle \cdot \rangle$ is taken over all indices $k$. (We will omit the explicit index $k$ in the following equations.)
The rms height measures the amplitude of height fluctuations on the topography, where the midline is defined as $h=0$. In addition to the height fluctuation, we can also quantify the amplitude of slopes,
\begin{equation}
    h^\prime_\text{rms} = \left\langle \left(\frac{D}{D x}  h(x)\right)^2 \right\rangle^{1/2},
\end{equation}
where $D/D x$ is a discrete derivative in the $x$-direction.

A common way to compute discrete derivatives on experimental data is to use a \emph{finite-differences} approximation. Finite-differences approximate the height $h(x)$ locally as a polynomial (a Taylor series expansion). The first derivative can then be computed as
\begin{equation}
    \frac{\partial}{\partial x} h(x) \approx \frac{D}{D x} h(x)
    = \frac{h(x+\Delta x) - h(x)}{\Delta x} .
    \label{eq:fdfirst}
\end{equation}
This expression is called the first-order right-differences scheme. We will use the symbol $D$ for the discrete derivatives, and the term ``order'' here refers to the truncation order, or how fast the error decays with grid spacing $\Delta x$: it drops linearly with decreasing $\Delta x$ in this scheme. Another interpretation is that the truncation order gives the highest exponent of the polynomial used to interpolate between the points $x$ and $x+\Delta x$. The derivative of a linear interpolation is constant between these points and given by Eq.~\eqref{eq:fdfirst}.

We can also quantify the amplitude of higher derivatives,
\begin{equation}
    h^{(\alpha)}_\text{rms} = \left\langle \left(\frac{D^\alpha}{D x^\alpha}  h(x)\right)^2 \right\rangle^{1/2},
    \label{eq:rmsalpha}
\end{equation}
where $\alpha=2$ yields the rms curvature. A discrete formulation of the second derivative is
\begin{equation}
    \frac{D^2}{D^2 x} h(x) = \frac{h(x+\Delta x) - 2h(x) + h(x-\Delta x)}{\Delta x^2}.
    \label{eq:fdsecond}
\end{equation}
This expression is called the second-order central-differences approximation. Again, this can be interpreted as fitting a second-order polynomial to the three points $x-\Delta x$, $x$ and $x+\Delta x$ and interpreting the (constant) second derivative of this polynomial as the approximate second derivative of the discrete set of data points.
The third derivative is given by
\begin{equation}
    \frac{D^3}{D^3 x} h(x) = \frac{h(x+2\Delta x) - 3h(x+\Delta x) + 3h(x) - h(x-\Delta x)}{\Delta x^3},
    \label{eq:fdthird}
\end{equation}
which again can be interpreted in terms of fitting a cubic polynomial to (four) collocation points.

\begin{figure}
    \centering
    \includegraphics[width=\columnwidth]{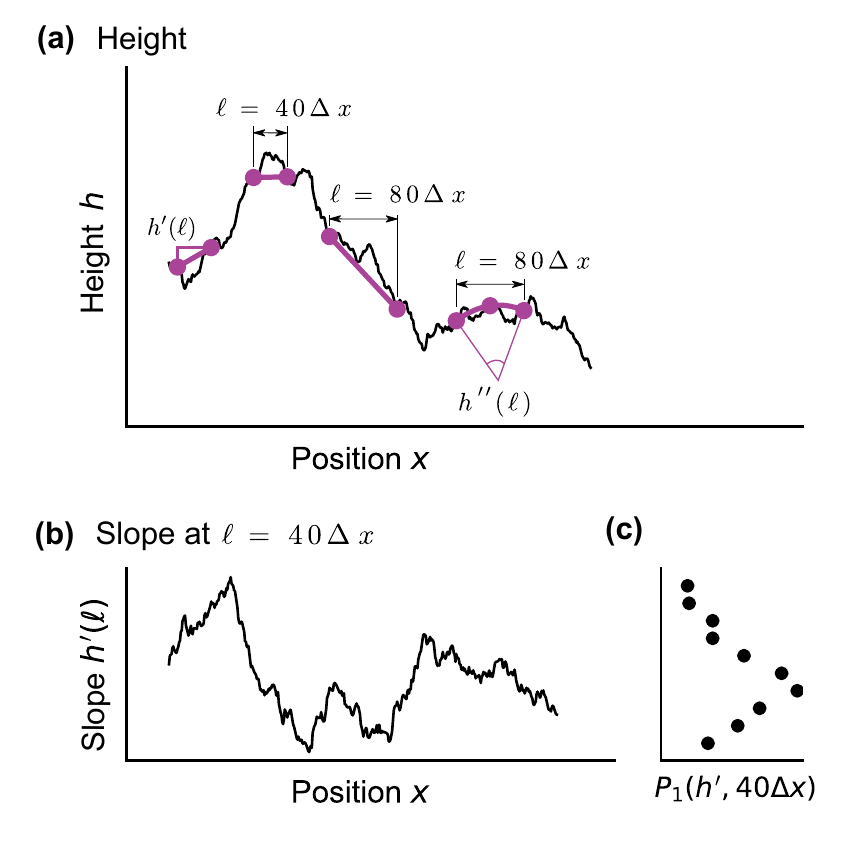}
    \caption{Illustration of the basic idea behind the scale-dependent roughness parameters. \textbf{(a)} Example line scan showing the computation of slopes $h^\prime(\ell)$ and curvatures $h^{\prime\prime}(\ell)$ from finite differences. A scale can be attached to this computation by computing these finite differences at different distances $\ell$, shown for $\ell=40\Delta x$ and $\ell = 80\Delta x$ where $\Delta x$ is the sample spacing. Similarly, the curvature at a finite scale $\ell$ is given by fitting a quadratic function through three points spaced a distance $\ell / 2$. \textbf{(b)} Local slope, obtained at a distance scale of $\ell=40 \Delta x$ for the line scan shown in panel (a). The slope is defined for each sample point since we can compute it for overlapping intervals. \textbf{(c)} Distribution of the local slope obtained from the slope profile shown in panel (b). The rms slope for this length scale is equal to the width of this distribution.
    }
    \label{fig:derivative}
\end{figure}

We can generally write the discrete derivative as a weighted sum of the collocation points $x_k$,
\begin{equation}
    \frac{D^\alpha}{D^\alpha x} h(x_k) = \frac{1}{\Delta x^\alpha} \sum_{l=-\infty}^\infty c_l^{(\alpha)} h(x_{k+l}).
    \label{eq:weightedsum}
\end{equation}
The values $c_l^{(\alpha)}$ are called the \emph{stencil} of the derivative operator and $\alpha$ indicates the order of the derivative. For the above derivatives,
\begin{align}
    c^{(1)}_{0} &= -1, c^{(1)}_1 = 1, \\
    c^{(2)}_0 &= -2, c^{(2)}_{\pm 1} = 1
    \quad\text{and}\\
    c^{(3)}_0 &= 3, c^{(3)}_{1} = -3, c^{(3)}_{-1} = -1, c^{(3)}_{2} = 1,
\end{align}
and all other $c^{(\alpha)}_l$s are zero. Higher-order derivatives lead to wider stencils.

\subsection{Computing scale-dependent roughness parameters in real space}
\label{sec:sdrp}

The discrete derivatives of the preceding section are all defined on the smallest possible scale that is given by the sample spacing $\Delta x$ and have an overall width of $\alpha\Delta x$. It is straightforward to attach an explicit scale to these derivatives, by evaluating Eq.~\eqref{eq:weightedsum} over a sample spacing $\eta\Delta x$ (with integer $\eta$) rather than $\Delta x$,
\begin{equation}
    \frac{D^\alpha_{(\eta)}}{D^\alpha_{(\eta)} x} h(x) \equiv
    \frac{1}{(\eta\Delta x)^\alpha} \sum_{l=-\infty}^\infty c_l^{(\alpha)} h(x_{k+\eta l}).
    \label{eq:weightedsumscaled}
\end{equation}
We will call the factor $\eta$ the \emph{scale-factor}. The corresponding derivative is measured at the \emph{distance scale} $\ell = \alpha \eta \Delta x$.

Figure~\ref{fig:derivative}a illustrates this concept. For a simple right-differences scheme as given by Eq.~\eqref{eq:fdfirst}, the scale-dependent first derivative is simply the slope of the two points at distance $\ell$. For the second derivative given by Eq.~\eqref{eq:fdsecond}, we fit a quadratic function through three points with overall spacing $\ell$ and the curvature of this function is the scale-dependent second derivative.

We now define the \emph{scale-dependent roughness parameter (SDRP)} as
\begin{equation}
    h^{(\alpha)}_\text{SDRP}(\alpha \eta \Delta x) = \left\langle \left(\frac{D^\alpha_{(\eta)}}{D^\alpha_{(\eta)} x} h(x)\right)^2 \right\rangle_{\text{domain}}^{1/2}.
    \label{eq:scaledepparareal}
\end{equation}
This new function defines a series of descriptors for the surface that are analogous to the rms slope ($h^{(1)}_\text{SDRP}\equiv h^{\prime}_\text{SDRP}$) and to the rms curvature ($h^{(2)}_\text{SDRP}\equiv h^{\prime\prime}_\text{SDRP}$); but instead of being a single scalar value, they represent curves as a function of the distance scale $\ell=\alpha \eta \Delta x$. 

The distance scale $\ell$ is only clearly defined for the stencils of lowest truncation order. For the $n$-th derivative, those can be interpreted as fitting a polynomial of order $n$ to $n+1$ data points (see Fig.~\ref{fig:derivative}a). The $n$-th derivative of this polynomial is then a constant over the width of the stencil; this width must then equal the distance scale $\ell$. Higher truncation orders can be interpreted as fitting a polynomial of order $m>n$ to $m+1$ data points. The $n$-th derivative is not constant over the stencil and it is not clear what the corresponding length scale is. We will here only use stencils of lowest truncation order where the distance scale is clear.

For non-periodic topographies we need to take care to only include derivatives that we can actually compute (i.e.\ where the stencil remains in the domain of the topography). This is indicated by the subscript ``domain'' in Eq.~\eqref{eq:scaledepparareal}.

\subsection{Beyond root-mean-square parameters: Computing the full distribution}
\label{sec:fulldistribution}

The rms value, such as the one defined in Eq.~\eqref{eq:scaledepparareal}, characterizes the amplitude of fluctuations, or the width of the underlying distribution function. Rather than looking at this single parameter, we can also determine the full scale-dependent distribution. Formally we can write this distribution as
\begin{equation}
    P_{\alpha}(\chi; \eta) = \left\langle \delta\left(\chi - \frac{D^\alpha_{(\eta)}}{D^\alpha_{(\eta)} x} h(x)\right)\right\rangle
\end{equation}
where $\delta(x)$ is the Dirac-$\delta$ function and $\chi$ the value of the derivative (of order $\alpha$) that we are interested in. In any practical (numerical) determination of the distribution, we broaden the $\delta$-function into individual bins and count the number of occurrences of a certain derivative value.

To illustrate this concept on the example of the slope ($\alpha=1$), Fig.~\ref{fig:derivative}b shows the scale-dependent derivative at $\ell=40\Delta x$ of the line scan shown in Fig.~\ref{fig:derivative}a. The distribution function of the slopes at this scale, $P_1(h', 40\Delta x)$, is then obtained by counting the occurence of a certain slope value. The resulting distribution is shown in Fig.~\ref{fig:derivative}c.

The rms parameters defined in the previous section are the second moments of this distribution,
\begin{equation}
    h^{(\alpha)}_\text{SDRP}(\alpha \eta \Delta x) = \left[ \int d\chi\, \chi^2 P_{\alpha}(\chi; \eta) \right]^{1/2}.
\end{equation}
The second moment characterizes the underlying distribution fully only if this distribution is Gaussian. We will see below that, for example, scanning probe artifacts introduce deviations from Gaussianity that we can easily detect once we have the full distribution function. In summary, the measurement of probability distributions as a function of distance scale for arbitrary derivatives (such as slope or curvature) enables the calculation of a generalized set of scale-dependent statistical parameters, including higher moments or the commonly used metrics skewness or kurtosis.

\section{Analysis: Relationship of scale-dependent roughness parameters to other methods}

\subsection{Relationship to the autocorrelation function}
\label{sec:acf}

A common way of analyzing the statistical properties of surface topography is the height-difference autocorrelation function (ACF) $A(\ell)$. (See Ref.~\onlinecite{wang_usefulness_2018} for an authoritative discussion of properties and use.) The ACF is defined as
\begin{equation}
\begin{split}
    A(\ell)
    &=
    \frac{1}{2}
    \left\langle \left[h(x+\ell) - h(x)\right]^2 \right\rangle
    \\
    &=
    \left\langle
    \frac{1}{2}
    h^2(x)
    +
    \frac{1}{2}
    h^2(x+\ell)
    -
    h(x)h(x+\ell) \right\rangle,
\end{split}
\label{eq:acf}
\end{equation}
which is commonly reduced to
\begin{equation}
    A(\ell)
    =
    h_\text{rms}^2
    -
    \left\langle
    h(x)h(x+\ell) \right\rangle.
    \label{eq:acfexplicitperiodic}
\end{equation}
The ACF has the limiting properties $A(0)=0$ and $A(\ell\to\infty)=h_\text{rms}^2$.

Equation~\eqref{eq:acf} resembles the finite-differences expression for the first derivative, Eq.~\eqref{eq:fdfirst}. Indeed, we can rewrite the ACF as
\begin{equation}
    A(\eta\Delta x)
    =
    \frac{1}{2}
    \left\langle \left[\frac{D_{(\eta)}}{D_{(\eta)} x} h(x)\right]^2 \right\rangle (\eta\Delta x)^2
    \label{eq:acf2}
\end{equation}
using the scale-dependent derivative. The scale-dependent rms slope then becomes
\begin{equation}
    h_\text{SDRP}^\prime(\ell) = \left[2A(\ell)\right]^{1/2}/\ell.
    \label{eq:hprime}
\end{equation}
The height-difference ACF can hence be used to compute the scale-dependent slope introduced above.

We now show that we can also express higher-order derivatives in terms of the ACF. Using the stencil of the second derivative given in Eq.~\eqref{eq:fdsecond}, the scale-dependent second derivative can be written as
\begin{equation}
    h^{\prime\prime}_\text{SDRP}(\ell)
    =
    \frac{4}{\ell^2}
    \left\langle
    \left[h(x+\ell/2) - 2h(x) + h(x-\ell/2)\right]^2
    \right\rangle^{1/2}.
\end{equation}
We can rewrite this expression as
\begin{equation}
\begin{split}
    h^{\prime\prime}_\text{SDRP}(\ell)
    =
    \frac{4}{\ell^2}
    \left\langle\right.
    &
    6h^2(x) - 8h(x)h(x+\ell/2) \\
    &  + 2 h(x) h(x+\ell)
    \left.\right\rangle^{1/2}.
\end{split}
\end{equation}
and use Eq.~\eqref{eq:acfexplicitperiodic} to introduce the ACF into this expression, yielding
\begin{equation}
    h_\text{SDRP}^{\prime\prime}(\ell)
    =
    4
    \left[8 A(\ell /2) - 2 A(\ell)\right]^{1/2}/\ell^2.
    \label{eq:scaledepcurvperiodic}
\end{equation}
Similarly, the scale-dependent third derivative from the stencil given in Eq.~\eqref{eq:fdthird} becomes
\begin{equation}
    h^{\prime\prime\prime}_\text{SDRP}(\ell)
    =
    \frac{27}{\ell^3}
    \left[
    30 A(\ell/3) - 12 A(2\ell/3) + 2 A(\ell)
    \right]^{1/2}.
\end{equation}
We can therefore relate the scale-dependent root-mean-square slope, curvature, or any other higher-order derivative to the ACF. Indeed, we will use these relationships to compute SDRPs in Section~\ref{sec:discussion}.

In summary, we have shown that the commonly used ACF function can be thought of a specific case of the SDRP analysis: being equivalent to the finite-differences calculation of scale-dependent slope. We further showed that the ACF function can be used as one method to compute higher-order SDRPs. 
 
\subsection{Relationship to the variable-bandwidth method}

We now introduce an alternative way to arrive at SDRPs based on a different notion of scale.
Notice that the 
discussion leading up to Eq.~\eqref{eq:scaledepparareal} does not involve the length $L$
of the line scan. This length is only relevant when it comes to determining an upper limit 
for the stencil length $\ell=\alpha \eta \Delta x$, which is the notion of scale in a measurement 
based on Eq.~\eqref{eq:scaledepparareal}. Alternatively, we could interpret $L$ as the relevant 
scale, and study scale-dependent roughness by varying $L$. This interpretation leads to a class
of methods which have been referred to as scaled windowed variance methods \cite{cannon_evaluating_1997} or variable bandwidth methods (VBMs).
They differ only in the way how the data is-detrended and are called the bridge method (attributed to Mandelbrot), roughness around the mean height (MHR) \cite{moreira_fractal_1994} (sometimes termed VBM~\cite{schmittbuhl_reliability_1995}), 
detrended fluctuation analysis (DFA) \cite{peng_mosaic_1994, peng_quantification_1995}, and
roughness around the rms straight line (SLR) \cite{moreira_fractal_1994}.

In all cases, one performs multiple roughness measurements on the same specimen (or the same material) but with different scan sizes $L$. Plotting the rms height $h_\text{rms}$ from these measurements versus scan size $L$, or the rms slope $h_\text{rms}^\prime$ versus scan resolution (the smallest measurable scale) yields insights into the multi-scale nature of surface topography. An example of an experimental realization of this idea is the classic paper by Sayles \& Thomas~\cite{sayles_surface_1978}. We have used and discussed this approach in the past to characterize the topography of diamond thin films~\cite{gujrati_combining_2018,gujrati_comprehensive_2021}. 

These methods can be generalized for the analysis of single measurements. Consider a line scan $h(x_k)$ of length $L$. The scan is partitioned into $\zeta\geq 1$ segments of length $\ell(\zeta)=L/\zeta$ (with $\ell \leq L$ now being the relevant scale). The dimensionless number $\zeta$, which we call the \emph{magnification}, defines the scale. Some
authors use sliding windows rather than exclusive segments \cite{moreira_fractal_1994,sandfeld_deformation_2014}.

The VBM considers the rms height fluctuations in each of the segments, i.e.\ one computes the standard deviation of the height $h_{\text{VBM},i}(\zeta)$ within segment $i$ at magnification $\zeta$, and then takes the average over all $i$ to compute a scale-dependent $h_\text{VBM}(\zeta)$. Some authors (including ourselves~\cite{hinkle_emergence_2020}) have tilt corrected the individual segments, i.e.\ each segment is detrended by subtracting the corresponding mean height and slope (obtained by linear 
regression of the data in the segment) before computing $h_{\text{VBM},i}(\zeta)$; this approach is called the DFA~\cite{peng_mosaic_1994,peng_quantification_1995} while without tilt correction it is called MHR. In the bridge method, the connecting line between the first and last point in each segment is used for detrending (see e.g. Ref.~\cite{schmittbuhl_reliability_1995}).

\begin{figure}
    \centering
    \includegraphics[width=\columnwidth]{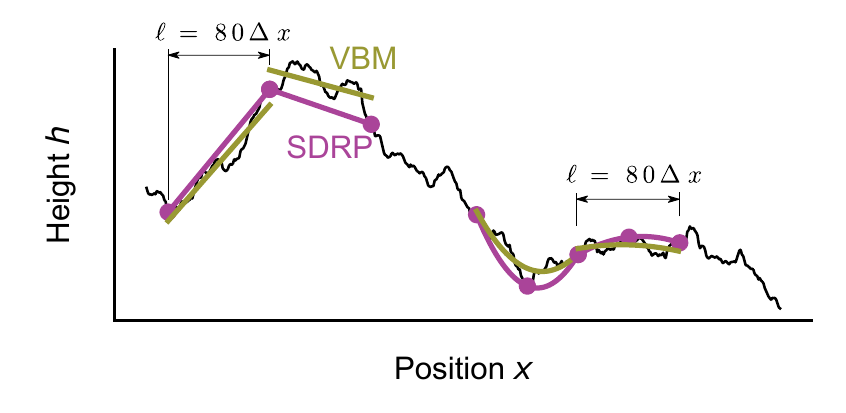}
    \caption{Illustration of the computation of scale-dependent roughness parameters from the variable bandwidth method (VBM). While in finite differences, the slope is computed between two point at distance $\ell$, in the VBM we fit a trend line to a segment of width $\ell$. Similarly for the second derivative, the finite-differences estimation fits a quadratic function through three points while in the VBM we fit a quadratic trend line through all data points in an interval of length $\ell$. }
    \label{fig:sdrpvsvbm}
\end{figure}

These VBMs are extremely similar to the SDRP. When computing the slope in the SDRP, we compute it by simply connecting the two boundary points at $x=i \ell(\zeta)$ and $x=(i+1) \ell(\zeta)$ with a straight line, identical to the bridge method. In the DFA, we use all data points between the two boundary points and fit a straight line using linear regression. Detrending can be generalized to higher order polynomials, but this has to the best of our knowlegde not been reported in the literature. The relationship between SDRP and VBMs with detrending of order $1$ and $2$ is conceptually illustrated in Fig.~\ref{fig:sdrpvsvbm}.

In the DFA, the slope of the trend line is simply used as a reference for the computation of fluctuations around it. We are not aware of any work that has analyzed the scaling of the slope with magnification. We here propose that this slope can be used to determine an alternative measure of the scale-dependent rms slope, $h_\text{VBM}^\prime(\zeta)$, obtained at magnification $\zeta$ or distance scale $\ell=L/\zeta$. $h_\text{VBM}^\prime(\zeta)$ is simply the standard deviation of slopes obtained within all segments $i$ at a certain magnification $\zeta$. We show in the examples below that this scale-dependent slope is virtually identical to the slope obtained from the SDRP.

We can use this idea to extend the DFA to higher-order derivatives. Rather that fitting a linear polynomial in each segment, we detrend using a higher-order polynomial. For extracting a scale-dependent rms curvature, we fit a second-order polynomial to the segment and interpret twice the coefficient of the quadratic term as the curvature. The standard deviation of this curvature over the segments then gives the scale-dependent second derivative, $h_\text{VBM}^{\prime\prime}(\zeta)$. Figure~\ref{fig:sdrpvsvbm} illustrates this concept, again in comparison to the SDRP that for the second-order derivative fits a quadratic function through just three collocation points.

An alternative route of thinking about VBMs is that they use a stencil whose number of coefficients equals the segment length. The stencil can be explicitly constructed from least squares regression (at each scale) of the polynomial coefficients. The closest equivalent to the SDRP would then be the respective VBM that uses sliding (rather than exclusive) segments. The difference to the SDRP is that the SDRP uses stencils of identical number of coefficients at each scale. In the examples provided below, we use a VBM that uses nonoverlapping segments.

In summary, we have shown that the various methods for computing scale-dependent height (such as VBM, DFA, and others) can be thought of as a special case of SDRP analysis: where the scale-dependent detrending only occurs for at most linear trend lines. We have then shown how those analyses can be extended to define a second method to compute SDRPs. 

\subsection{Relationship to the power spectral density}

Finally we outline a third way to arrive at SDRPs using the power spectral density (PSD), another common tool for the statistical analysis of topographies. Underlying the PSD is a Fourier spectral analysis, which approximates the topography map as the series expansion
\begin{equation}
    h(x) = \sum_n a_n \phi_n(x),
    \label{eq:seriesexpansion}
\end{equation}
where $\phi_n(x)$ are called \emph{basis functions}. The Fourier basis is given by
\begin{equation}
    \phi_n(x) = \exp(i q_n x),
    \label{eq:Fourierbasis}
\end{equation}
with $q_n = 2\pi n/L$ where $L$ is the lateral length of the sample. The inverse of Eq.~\eqref{eq:seriesexpansion} gives the expansion coefficients $a_n$ which are typically computed using a fast Fourier-transform algorithm. The PSD is then obtained as~\cite{jacobs_quantitative_2017}
\begin{equation}
    C^\text{1D}(q_n)=L|a_n|^2.
\end{equation}
Fourier spectral analysis is useful because a notion of scale is embedded in the definition Eq.~\eqref{eq:Fourierbasis}: The wavevectors $q_n$ describe plane waves with \emph{wavelength} $\lambda_n = 2\pi / q_n$.

This basis leads to spectral analysis of surface topography and derivatives are straightforwardly computed from the derivatives of the basis functions,
\begin{align}
    \frac{\partial}{\partial x} \phi_n(x) &= i q_n \phi_n(x) \quad\text{and}
    \label{eq:fourierfirst}
    \\
    \frac{\partial^2}{\partial^2 x} \phi_n(x) &= -q_n^2 \phi_n(x).
    \label{eq:fouriersecond}
\end{align}
We can write the Fourier-derivative generally as
\begin{equation}
    \frac{\partial^{\alpha}}{\partial x^{\alpha}} \phi_n(x) = \mathcal{D}_\alpha(q_n) \phi_n(x)
\end{equation}
with $\mathcal{D}_1(q_n)=i q_n$ for the first derivative and $\mathcal{D}_2(q_n)=-q_n^2$ for the second derivative. The $\mathcal{D}_\alpha(q_n)$ are complex numbers that we will call the \emph{derivative coefficients}.

The rms amplitude of fluctuations can be obtained in the Fourier picture from Parseval's theorem, that turns the real-space average in Eq.~\eqref{eq:rmsalpha} into a sum over wavevectors,
\begin{equation}
    h^{(\alpha)}_\text{rms} = \left[ \sum_n \left|\mathcal{D}_{\alpha}(q_n) a_n\right|^2 \right]^{1/2}.
    \label{eq:fourierrmsalpha}
\end{equation}

The notion of a scale-dependence can be introduced in the Fourier picture by removing the contribution of all wavevectors $|q_n|>q_c$ larger than some characteristic wavevector $q_c$, i.e.\ setting the corresponding expansion coefficients $a_n$ to zero. This means there are no longer short wavelength contributions to the topography. We will refer to this process as \emph{Fourier filtering}.
Fourier filtering can be used to introduce a scale-dependent roughness parameter, e.g.
\begin{equation}
    h^{(\alpha)}_\text{PSD}(q_c) = \left[\sum_n \left|\mathcal{D}^\text{F}_{\alpha}(q_n;q_c)\right|^2 C^\text{1D}(q_n) \right]^{1/2}
    \label{eq:psdalpha}
\end{equation}
with $\mathcal{D}^\text{F}_{\alpha}(q_n;q_c)=\Theta(q_c - |q_n|)\mathcal{D}_{\alpha}(q_n)$ that we call the Fourier-filtered derivative and $\Theta(x)$ is the Heaviside step function. Note that we have expressed Eq.~\eqref{eq:psdalpha} in terms of the PSD, which is typically obtained using a windowed topography if the underlying data is nonperiodic. In the examples that we show in Sec.~\ref{sec:discussion}, we applied a Hann window before computing the PSD and the scale-dependent derivatives.

We now show that Fourier-filtering and finite-differences are related concepts. We first interpret the finite-differences scheme in terms of a Fourier analysis. We apply the finite differences operation to the Fourier basis Eq.~\eqref{eq:Fourierbasis}. This yields
\begin{equation}
    \frac{D^\alpha_{(\eta)}}{D^\alpha_{(\eta)} x} \phi_n(x_k) =  \mathcal{D}_\alpha^\text{s}(q_n;\eta) \phi_n(x_k)
    \label{eq:fourierderivative}
\end{equation}
with
\begin{equation}
   \mathcal{D}_\alpha^\text{s}(q_n;\eta) = \frac{1}{(\eta\Delta x)^\alpha} \sum_{l=-\infty}^\infty c_l^{(\alpha)} \exp(i q_n \eta l\Delta x).
\end{equation}
Note that the right hand side of Eq.~\eqref{eq:fourierderivative} is fully algebraic, i.e.\ it no longer contains derivative operators, and the $\mathcal{D}_\alpha^\text{s}(q_n;\eta)$ are (complex) numbers. Inserting these derivative coefficients into Eq.~\eqref{eq:psdalpha} yields Eq.~\eqref{eq:scaledepparareal}. We have therefore unified the description of (scale-dependent) derivatives in the Fourier basis and finite-differences in terms of the derivative coefficients $\mathcal{D}_\alpha$.

\begin{figure}
    \centering
    \includegraphics{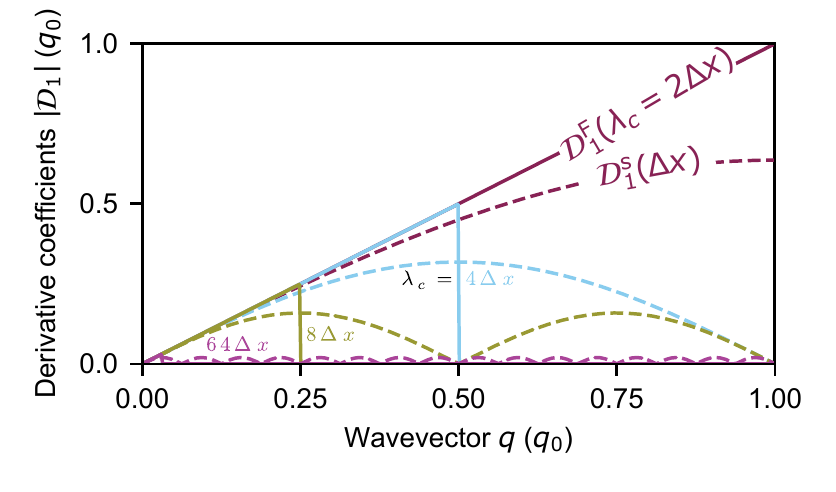}
    \caption{
    Derivative coefficients for finite differences $\mathcal{D}_1^\text{s}$ and the Fourier-filtered derivative $\mathcal{D}_1^\text{F}$ for different distance scales $\ell$. The coefficients agree at small wavevectors $q$. The maximum of the coefficient agrees if the filter wavelength $\lambda_c=2\ell$, corresponding to the Nyquist sampling theorem (see text).} 
    \label{fig:derivative_coefficients}
\end{figure}

The remaining question is how the scale $\ell$ used to compute the finite-differences relates to the wavevector $q_c$ used in Fourier-filtering. 
Figure~\ref{fig:derivative_coefficients} shows $\mathcal{D}_1^\text{F}(\lambda_c)$ and $D_1^\text{s}(\ell)$ for different values of $\ell$ and $\lambda_c$. The location of the maximum of these derivative coefficients agrees if $\eta\Delta x\equiv \ell/\alpha =\lambda_c/2=\pi/q_c$. For first derivatives ($\alpha=1$), $\ell=\Delta x$. This is the Nyquist sampling theorem, which states that the shortest wavelength we can resolve is $\lambda_c=2\Delta x$. This means to compare SDRP, VBM and PSD, we need to choose a filter cutoff of $q_c=\alpha \pi/\ell$ in the latter. Note that in the SDRP, the (soft) cutoff emerges implicitly from the finite-difference formulation.

In summary, we have shown that the scale-dependent calculations, which were defined in real-space in Sec.~\ref{sec:sdrp}, can be equivalently computed in frequency-space using the PSD. The results should be the same, however, frequency-space calculations have the shortcomings that nonperiodic topographies need to be windowed, and a filter cutoff needs to be applied.  

\section{Discussion: Application of scale-dependent roughness parameters and advantages over other methods}
\label{sec:discussion}

\subsection{Application to a synthetic self-affine surface}

We first apply the concepts presented above to a synthetic self-affine topography. The topography has been first presented in Ref.~\onlinecite{jacobs_quantitative_2017} and consists of three virtual ``measurements'' of a large ($65\,536\times 65\,536$ pixels) self-affine topography generated with a Fourier-filtering algorithm. (See Refs.~\onlinecite{ramisetti_autocorrelation_2011,jacobs_quantitative_2017} for more information on this algorithm.) The topography has a Hurst exponent of $H=0.8$, a pixel size (resolution) of $\Delta x = \Delta y = 2$~nm and a short wavelength cutoff $\lambda_s \approx 10$~nm. The (Fourier-space) power is zero below this cutoff. For wavevectors with nonzero power, we choose a random phase and an amplitude from a Gaussian distribution with a variance given by the PSD. This surface was subsampled in three block of $500\times 500$ pixels at overall lateral sizes of $100~\text{µm}\times 100~\text{µm}$, $10~\text{µm}\times 10~\text{µm}$ and $1~\text{µm}\times 1~\text{µm}$ to emulate measurement at different resolution. Each of these virtual measurements is independently tilt-corrected. The data for the three subsampled topographies is available online~\cite{self-affine_synthetic_surface_v1}.

\begin{figure}
    \centering
    \includegraphics[width=\columnwidth]{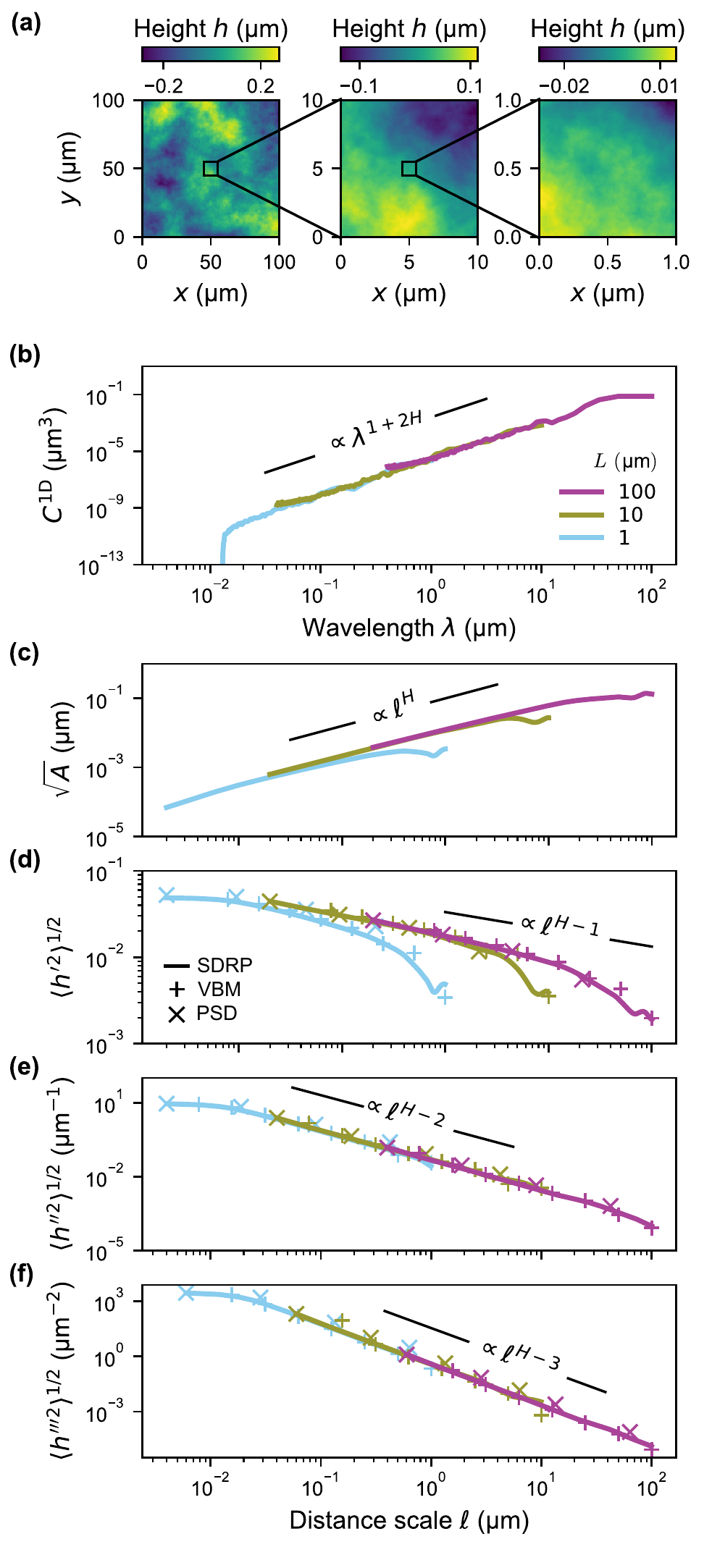}
    \caption{
    Example of scale dependent-roughness parameters for an ideal self-affine surface with Hurst exponent $H=0.8$. \textbf{(a)} A large surface was subsampled in three topographies of $500\times 500$ pixels at different resolution. \textbf{(b)} Individual PSDs displayed as a function of wavelength $\lambda=2\pi / q$, where $q$ is the wavevector. \textbf{(c)} Square-root of the ACF displayed as a function of distance scale $\ell$. \textbf{(d)} Scale-dependent rms slope. \textbf{(e)} Scale-dependent rms curvature. \textbf{(f)} Third derivative as an example of how this method can be used to go beyond traditional analysis. Color indicates the topography. The figure shows results from the three SDRPS (ACF, VBM and PSD), showing that the results agree. The solid black lines in panels (b-f) shows the power-law scaling of an ideal self-affine topography with Hurst exponent $H$. Note that the deviations from power-law scaling at large scales are especially visible in panels (c) and (d) because of the smaller range of values on the y-axis.}
    \label{fig:self_affine_scale_dependent_parameters}
\end{figure}

Figure~\ref{fig:self_affine_scale_dependent_parameters}a shows the topography map of these three emulated measurements. The measurements zoom subsequently into the center of the topography. The one-dimensional PSDs ($C^\text{1D}$, Fig.~\ref{fig:self_affine_scale_dependent_parameters}b) of the three topographies align well, showing zero power below the cutoff wavelength of $\lambda_s$. Note that unlike most authors (with few exceptions in geophysics~\cite{sagy_evolution_2007,candela_roughness_2012,brodsky_constraints_2016,candela_minimum_2016,thom_nanoscale_2017}) or even our own prior work, we display the PSD as a function of wavelength $\lambda = 2\pi / q$ where $q$ is the wavevector; this facilitates comparison with the real space techniques introduced above, and also wavelengths are more intuitively understandable than wavevectors.  Since the topography is self-affine, the PSD scales as $C^\text{1D}\propto \lambda^{1+2H}$ as indicated by the solid line.

The ACF (or rather its square-root) is shown in Fig.~\ref{fig:self_affine_scale_dependent_parameters}c. The ACF and all other scale-dependent quantities reported below are obtained from averages over adjacent line scans, i.e.\ from profiles and not the two-dimensional topography. This is compatible with how $C^\text{1D}$ is computed (see Ref.~\cite{jacobs_quantitative_2017}). The ACFs from the three measurements line up and follow $\sqrt{A}\propto \ell^{H}$ (see solid black line in Fig.~\ref{fig:self_affine_scale_dependent_parameters}c). Note that the ACF does not drop to zero for $\ell < \ell_s \equiv \lambda_s/2$ as the PSD did. This behavior becomes clearer by inspecting the scale-dependent slope $h^\prime_\text{SDRP}(\ell)=\sqrt{2A(\ell)}/\ell$ that saturates at a constant value for $\ell<\ell_s$. This is the true rms slope that is computed when all scales are considered. For large $\ell$, the rms slope scales as $h^\prime_\text{SDRP}\propto \ell^{H-1}$ (solid black line in Fig.~\ref{fig:self_affine_scale_dependent_parameters}d). We would also like to point out that ACF and rms slope of the individual measurements appear to line up worse than the PSD, but this is simply because the range of values for ACF and rms slope is much narrower than for the PSD.

Finally, we display the scale-dependent curvature $h^{\prime\prime}_\text{SDRP}(\ell)$ in Fig.~\ref{fig:self_affine_scale_dependent_parameters}e. Like the rms slope, the curvature saturates for $\ell<\ell_s$ to the ``true'' small-scale value of the curvature. The curvatures of the three individual measurements again line up and follow $h^{\prime\prime}_\text{SDRP}(\ell)\propto\ell^{H-2}$ because of the self-affine character of the overall surface. We would like to point out that the rms curvature shown here has been computed from Eq.~\eqref{eq:scaledepcurvperiodic} that is strictly only applicable to periodic topographies. However, we expect insignificant errors except at large distance scales. These errors manifest themselves in negative values for $h^{\prime\prime}_\text{SDRP}$, that violate the fact that the rms curvature needs to be strictly positive. We do not show the result of the calculation for distances $\ell$ larger then the first distance where negative values occur.

In our derivation above we have presented alternative routes for obtainig scale-dependent parameters from the VBM and PSD. The plusses ($+$) in Figs.~\ref{fig:self_affine_scale_dependent_parameters}d and e show the rms slope and curvature obtained using the VBM, while the crosses show the results obtained using the PSD. They align well with the respective parameters obtained from the ACF and only deviate at large scales. In summary, all three routes (ACF, VBM, PSD) for obtaining SDRPs are equally valid and lead to compatible results. The advantage of the ACF and the VBM is that they are directly (without windowing) applicable to nonperiodic data.

We have now demonstrated three independent ways of obtaining scale-dependent slopes, curvatures and higher-order derivatives. We would like to point out that all three routes constitute novel uses of the underlying analysis methodology. Our primary tool in what follows will be the SDRP; however, we have demonstrated that the SDRP, VBM and PSD yield equivalent results. The broader importance of using scale-dependent slopes and curvatures over the ``bare''  ACF, VBM or PSD is that it is straightforward to interpret the meaning of these parameters. We all have (geometric) ideas of slopes and curvatures but it is difficult to attribute a geometric meaning to a value of the PSD (that can even differ in unit, see discussion in Ref.~\onlinecite{jacobs_quantitative_2017}).

\subsection{Detecting tip artifacts in simulated topography measurements}

We now turn to another example, the analysis of tip artifacts. This will exploit a power of the SDRP, namely the fact that we can compute the full underlying distribution of arbitrary derivatives outlined in Section~\ref{sec:fulldistribution}.

\begin{figure*}
    \centering
    \includegraphics[width=\textwidth]{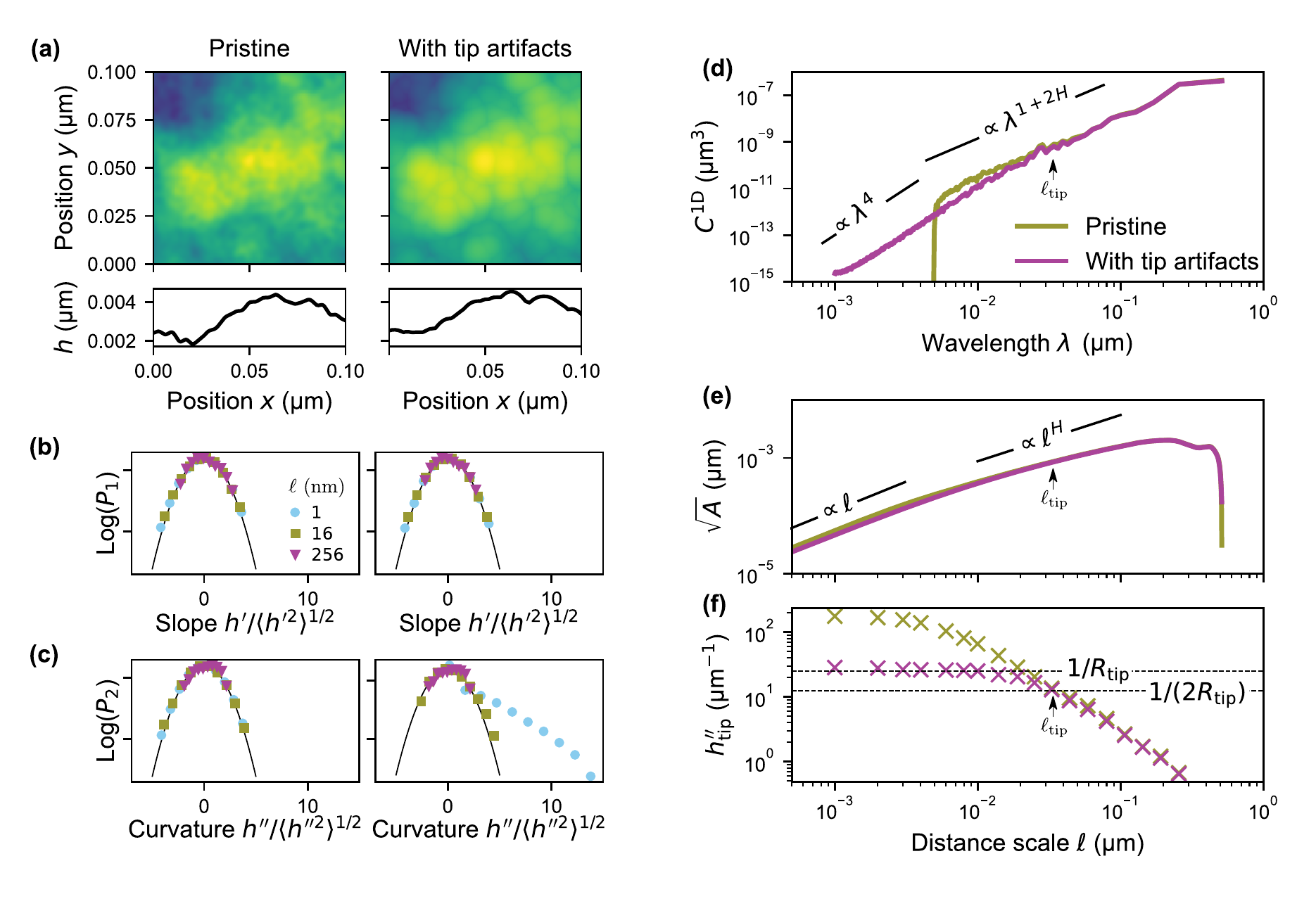}
    \caption{
    Scale-dependent roughness parameters for the analysis of tip artifacts. \textbf{(a)} A computer-generated ``pristine'' topography was scanned with a virtual tip of $R_\text{tip}=40$~nm radius. The bottom row shows cross-sectional profiles of of the maps shown above. Maps and profile show clear blunting of the peaks and cusps in the valleys (see text for more discussion). (b) Distribution of slopes at distance scales $\ell=1$~nm (circles $\bullet$), $16$~nm (squares $\blacksquare$) and $256$~nm (triangles $\blacktriangledown$). (c) Distribution of curvatures at these scales. Both slopes and curvatures are obtained in the $x$-direction. The bottom set of data points shows the pristine topography, the top set the topography with tip artifacts. Black solid lines show the normal distribution. (d) PSDs and (e) ACFs of both topographies. 
    (f) Minimum curvature $h_{\text{tip}}^{\prime\prime}$ (see text).
    } 
    \label{fig:tip_artefact_synthetic}
\end{figure*}

Figure~\ref{fig:tip_artefact_synthetic}a shows $0.1~\text{µm}\times 0.1~\text{µm}$ sections of two (periodic) topographies of total size $0.5~\text{µm}\times 0.5~\text{µm}$. The first topography is pristine and was generated using the Fourier-filtering algorithm mentioned above. The second topography contains tip artifacts and was obtained from the pristine surface using the following nonlinear procedure: For every location $(x_i,y_i)$ on the topography we lower a sphere with radius $R_\text{tip}$ (here $40$~nm) towards a position $(x_i,y_i,z_i)$ until the sphere touches the pristine topography anywhere. The resulting $z$-position $z_i$ of the sphere is then taken as the ``measured'' height of the topography. This topography was discussed in Ref.~\onlinecite{jacobs_quantitative_2017} and the data files are available at Ref.~\onlinecite{synthetic_self-affine_topography_scanned_with_a_virtual_spherical_probe_v1}. The two curves underneath the maps in Fig.~\ref{fig:tip_artefact_synthetic}a are cross-sections through the middle of the respective topography.

It is clear from simply looking at the data in Fig.~\ref{fig:tip_artefact_synthetic}a that the scanning probe smoothens the peaks of the topography. Indeed the curvature near the peaks must be equal to $-1/R_\text{tip}$. Conversely, the valleys look like cusps that originate from the overlap of two spherical bodies. These cusps are sharp and should lead to large (in theory unbounded, but in practice bounded by resolution and noise) positive values of the curvature. Church \& Takacs~\cite{church_instrumental_1989,church_effects_1991} have pointed out that tip artifacts should lead to PSD $C^\text{1D}(q)\propto q^{-4}$, which is precisely a result of the cusps in the topography. (The Fourier transform of a triangle $\propto q^{-2}$, leading to a PSD $\propto q^{-4}$.) We have demonstrated in Refs.~\onlinecite{jacobs_quantitative_2017,gujrati_comprehensive_2021} numerically that this is indeed the case.

We are now in a position to more precisely look at the effect of tip radius. 
Figure~\ref{fig:tip_artefact_synthetic}b shows the scale-dependent slope distribution $P_1(h^\prime, \ell)$, normalized by the rms slope at the respective scale. The black solid line shows a Gaussian distribution (of unit width) for reference. It is clear that both our pristine topography (left columns) and the topography with tip radius artifacts (right column) follow a Gaussian distribution for the scale dependent slopes across scales from $1$~nm to $256$~nm shown in the figure.

The situation is different for the scale-dependent curvature, shown in Fig.~\ref{fig:tip_artefact_synthetic}c. While the pristine surface (left column) follows a Gaussian distribution, the topography with tip radius artifacts is only Gaussian for larger scales ($\ell=16$~nm and $256$~nm). There is a clear deviation at the smallest scales, showing an exponential distribution for positive curvature values, corroborrating the empirical discussion above that cusps leads to large positive values for the curvature. As argued above and in Refs.~\cite{church_instrumental_1989,church_effects_1991}, these cusps lead to a PSD $\propto q^{-4}\propto \lambda^4$. The PSDs of both topographies are shown in Fig.~\ref{fig:tip_artefact_synthetic}d, clearly showing that the artifacted surface crosses over to $C^\text{1D}\propto \lambda^4$ at a wavelength of $\lambda\sim 20-40$~nm, around half of $R_\text{tip}$ (see also discussion in Ref.~\onlinecite{jacobs_quantitative_2017}).

We note this cross-over to $\lambda^{4}$ is subtle and difficult to detect in measured data. Other measures, such as the ACF shown in Fig.~\ref{fig:tip_artefact_synthetic}e are unsuitable to detect these artifacts. The region where the $C^\text{1D}\propto \lambda^4$ shows up as a linear region in the square root of the ACF, $\sqrt{A}\propto \ell$. The exponent of $1$ from that region is too close to the exponent of $H=0.8$ to be clearly distinguishable.

We now propose an additional metric that is intended to more accurately detect the onset of the tip-radius artifact. Rather than looking at some measure for the full width of the distribution like our rms measures, we now ask the question of what is the minimum curvature value found at a specific scale $\ell$. We therefore evaluate
\begin{equation}
    h_\text{tip}^{\prime\prime}(\ell) = -\min_k \left[\frac{D^2_{(\ell)}}{D^2_{(\ell)}} h(x_k)\right].
    \label{eq:htipinfty}
\end{equation}
The crosses in Fig.~\ref{fig:tip_artefact_synthetic}f show this quantity for the pristine and the artifacted surface. It is clear that at small scales the curvature of the pristine surface is larger than the artifacted one. Additionally, the artifacted surface settles to $h_\text{tip}^{\prime\prime}(\ell)\approx 1/R_\text{tip}$ as $\ell\to 0$. This is a clear indicator that the curvature of the peaks on the artifacted surface is given by the tip radius.

In order to estimate a scale below which the data is unreliable, we search for the characteristic scale $\ell_\text{tip}$ where $h_\text{tip}^{\prime\prime}(\ell_\text{tip})= c/R_\text{tip}$ with an empirical factor $c$. Figure~\ref{fig:tip_artefact_synthetic}f shows this condition as a dashed horizontal line. We also indicate the scale $\ell_\text{tip}$ in the ACF (Fig.~\ref{fig:tip_artefact_synthetic}e) and in the PSD. The factor $c$ was chosen such that $\lambda_\text{tip}$ marks the crossover from artifacted $C^\text{1D}\propto \lambda^4$ to the pristine $C^\text{1D}\propto \lambda^{1+2H}$. We will use the same factor when analyzing experimental data in the next section.

We note that for experimental data there is no ``pristine'' measurement available for comparison. Our proposed measure is useful because it can be robustly and automatically carried out on large sets of measurements; by contrast, the detection of $C^\text{1D}\propto \lambda^4$ is difficult because fitting exponents requires data over at least a decade in length and carries large errors~\cite{clauset_power-law_2009}.

\subsection{Application to an experimental measurement}

\begin{figure}
    \centering
    \includegraphics[width=\columnwidth]{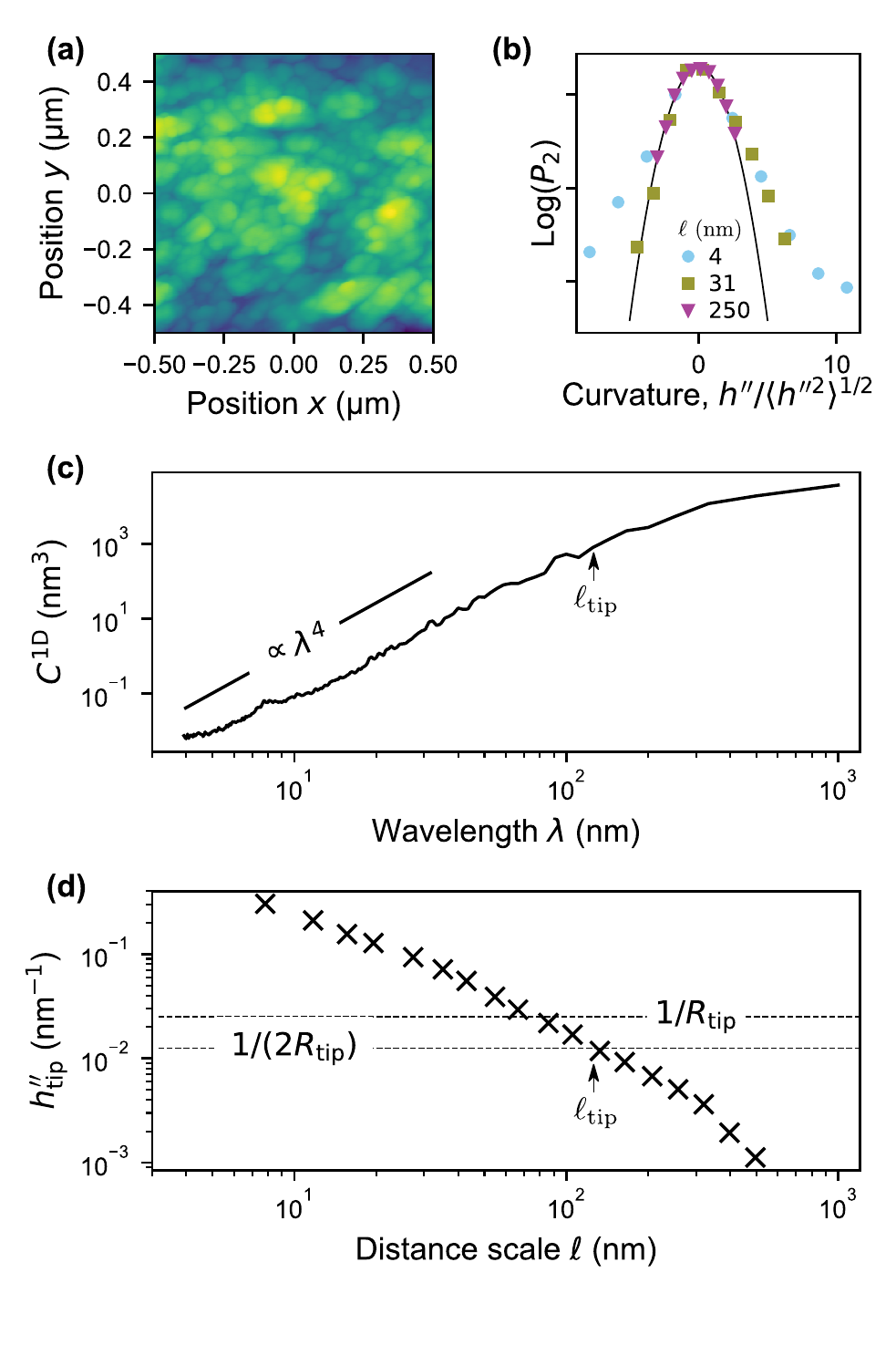}
    \caption{
    AFM measurement of an ultrananocrystalline diamond film. (a) AFM measurement showing the smoothing of peaks similar to the emulated scans shown in Fig.~\ref{fig:tip_artefact_synthetic}a. (b) Normalized curvature distribution at distance scales $\ell=4$~nm (circles $\bullet$), $31$~nm (squares $\blacksquare$) and $250$~nm (triangles $\blacktriangledown$).
    (c) power spectral density (PSD) of the measurement. The black solid line shows scaling with $\lambda^4$ that indicates tip artifacts. (d) Peak curvature $h_{\text{tip},\alpha}^{\prime\prime}$ (see text) used for estimation tip radius artifacts.  }
    \label{fig:tip_artefact_uncd}
\end{figure}

As a final example, we turn to an atomic force microscopy (AFM) scan of an ultrananocrystalline diamond (UNCD) film. The film has been described in detail in Ref.~\onlinecite{gujrati_combining_2018} and the underlying dataset is accessible in Ref.~\onlinecite{ultrananocrystalline_diamond_(uncd)_v1}. Figure~\ref{fig:tip_artefact_uncd}a shows the AFM scan. The peaks have rounded tips similar to the synthetic scan shown in Fig.~\ref{fig:tip_artefact_synthetic}a. The curvature distribution (Fig.~\ref{fig:tip_artefact_uncd}b) also has a similar characteristic to our synthetic topography (Fig.~\ref{fig:tip_artefact_synthetic}c). At large scales, the distribution is approximately Gaussian (shown by the solid black line) with a slight tilt towards larger curvature values. We can only speculate on the origin of these deviations, but believe that they must be related to the preferential orientation of how these film grow in vapor deposition processes. At smaller scales, we see deviations to higher curvature values, indicative of the cusps that are characteristic of tip artifacts.

Unlike the synthetic surfaces, the scale-dependent tip curvature $h_\text{tip}^{\prime\prime}(\ell)$ (Fig.~\ref{fig:tip_artefact_synthetic}f) does not saturate to a specific value at small distances $\ell$. This is likely due to additional instrumental noise that contributes to small scale features of the data. Nevertheless, for $R_\text{tip}=40$~nm we can identify the region where $h_\text{tip}^{\prime\prime}(\ell)>1/(2 R_\text{tip})$ as unreliable, leading to a lateral length-scale of around $\ell_\text{tip} \approx 100$~nm below which the data is no longer reliable. The PSD (Fig.~\ref{fig:tip_artefact_uncd}c) shows $\lambda^4$ scaling below the characteristic wavelength $\ell_\text{tip}$.

\section{Summary \& Conclusions}

First, we demonstrated the calculation of scale-dependent parameters using a finite-differences scheme, with a variable distance scale. Then we showed that the commonly used autocorrelation function (ACF) can be interpreted as the scale-dependent root-mean-square (rms) finite-difference slope. This leads to a straightforward generalization of the ACF for higher derivatives, yielding for example a scale-dependent rms curvature. We have termed this characterization of first- and higher-order derivatives the \emph{scale-dependent roughness parameter (SDRP)} analysis. We have further generalized this analysis to estimate the curvature of peaks on the topography maps, an analysis that can be used to identify tip-radius artifacts. We also demonstrated the equivalence of computing scale-dependent roughness parameters from other conventional techniques: the variable bandwidth method and the power spectral density.

In summary, we proposed the use of a new SDRP analysis, and performed novel analysis to show how this is a generalization of commonly used roughness metrics. We suggest that this SDRP approach serves to harmonize competing roughness descriptors, but also offers advantages over those other methods, especially in terms of ease of calculation, intuitive interpretability, and detection of artifacts.
 
 \section*{Acknowledgements}

LP thanks the Deutsche Forschungsgemeinschaft (project EXC-2193/1 -- 390951807) and the European Research Council (StG-757343) for funding this work. TDBJ acknowledges funding from the U.S. National Science Foundation under award number CMMI-1727378.

\appendix

\section{Generalization to two dimensions}
\label{app:2d}

We here briefly outline the generalization of the SDRP to two-dimensions. The main difference is that in two dimensions the derivative becomes the (discrete) gradient $(D h/Dx, Dh /Dy)$, the curvature becomes the Hessian $(D^2h/Dx^2, D^2h/Dy^2, D^2h/Dx Dy)$ and higher-order derivatives contain additional cross terms. All averages are carried out over areas, not line scans. We can then for example define a scale-dependent \emph{gradient} as
\begin{equation}
    h^{\prime,\text{2D}}_\text{SDRP}(\ell) = \left\langle \left(\frac{D_{(\ell/\Delta x)}}{D_{(\ell/\Delta x)} x} h(x,y)\right)^2 +  \left(\frac{D_{(\ell/\Delta y)}}{D_{(\ell/\Delta y)} y} h(x,y)\right)^2 \right\rangle^{1/2},
\end{equation}
where the average $\langle\cdot\rangle$ now runs over the area. We note that in two-dimensions the situation may arise, where the scale factors $\eta_x = \ell/\Delta x$ and $\eta_y = \ell/\Delta y$ are no longer integer; this in particular happens if the aspect ratio of the individual pixel is not unity, $\Delta x\not = \Delta y$. In this case the additional (numerical) complexity arises, that one needs to interpolate between data points to measure the derivatives at the same distance scale in $x$- and $y$-direction.


%

\end{document}